# Dynamic ordering transitions in charged solid

Jian Sun[1†], Jiasen Niu[1†], Yifan Li[1], Yang Liu[1], L. N. Pfeiffer[2], K. W. West[2], Pengjie Wang[3]*, Xi Lin[1,4,5]*

[1]International Center for Quantum Materials, Peking University, Beijing 100871, China

[2]Department of Electrical Engineering, Princeton University, Princeton, New Jersey 08544, USA

[3]Department of Physics, Princeton University, Princeton, New Jersey 08544, USA

[4]Beijing Academy of Quantum Information Sciences, Beijing 100193, China

[5]CAS Center for Excellence in Topological Quantum Computation, University of Chinese Academy of Sciences, Beijing 100190, China

[†]These authors contributed equally to this work.

*xilin@pku.edu.cn; *pengjie.wang@princeton.edu

**Abstract**

The phenomenon of group motion is common in nature, ranging from the schools of fish, birds and insects, to avalanches, landslides and sand drift. If we treat objects as collectively moving particles, such phenomena can be studied from a physical point of view, and the research on many-body systems has proved that marvelous effects can arise from the simplest individuals. The motion of numerous individuals presents different dynamic phases related to the ordering of the system. However, it is usually difficult to study the dynamic ordering and their transitions through experiments. Electron bubble states formed in a two-dimensional electron gas, as a type of electron solids, can be driven by an external electric field and provide a platform to study the dynamic collective behaviors. Here, we demonstrate that noise spectrum is a powerful method to investigate the dynamics of bubble states. We observed not only the phenomena from dynamically ordered and disordered structures, but also unexpected alternations between them. Our results show that a dissipative system can convert between chaotic structures and ordered structures when tuning global parameters, which is concealed in conventional transport measurements of resistance or conductance. Moreover, charging the objects to study electrical noise spectrum in collective motions can be an additional approach to revealing dynamic ordering transitions.

**Keywords**

two-dimensional system, electron solid, dynamic ordering transition, noise spectrum measurement

## 1. Introduction

Collective motions are ubiquitous in creature behaviors and our daily life, and physics has turned out to be a unique tool to probe the complex consequences from simple units [1]. A typical example is the granular system [2, 3], which consists of discrete macroscopic solid grains and exhibits extraordinary phenomena, such as packing like solids while flowing like liquids [3]. Dynamic study of the granular system usually focuses on the spatial distributions and mechanical properties, where optical methods [4-6] (such as CCD cameras and X-ray tomography) and mechanical probes [5, 7] (such as accelerometers and pressure sensors) are applied. However, the dynamic motions of simple units are far from being understood [2, 3].

When a group of moving units is charged, it can be further probed electrically. Electron is probably the most well-understood charged object, so we might use condensed-matter systems based on electrons to study collective motion. Typical examples include multi-electron bubbles forming on the helium surface [8, 9] and electron solids such as Wigner crystals [10] and bubble states [11-14] forming in two-dimensional electron gas (2DEG), which are all composed of charged units with spatial periodicity. Considering the convenience of performing electrical measurements, electron solids forming in 2DEG serve as an appropriate choice. The charged units in a bubble state are electron bubbles and in a Wigner crystal are electrons. When a strong external electric field is applied, localized electron solids will be collectively depinned from their localized sites, offering examples of collective motions based on charged units [15-22]. Although the shape of the charged units may change during movement, the number of electrons in each charged unit remains constant for a particular state [11-14]. The motion of the depinned charged units, when interacting with disorder, will form different dynamic phases [23-27], which can be generally classified into two types according to the ordering: dynamically ordered phases (e.g., moving smectic phase [24]) and dynamically disordered phases (e.g., plastic flow phase [25, 27]). Moreover, dynamic ordering transitions are expected to happen [23], yet lack direct experimental evidence. Such a behavior is reminiscent of the pattern formation behavior in dynamic granular systems [28-30] as a non-equilibrium physics problem [31].

In this work, we distinguish the dynamic structures with contrasting ordering in a 2DEG system by noise spectrum measurements. Our ultra-high quality GaAs 2DEG sample (see Methods and Supplemental Material Section 1 for details) exhibits well-developed bubble states under a magnetic field, which serve as ideal platforms to investigate the dynamics of charged solids. From the dependence of the noise signals on the driving electric field, we observe unexpected alternations between dynamically ordered and disordered structures.

## 2. Results

The conductance $G$ of charged units, by measuring the differential value $dI/dV$, is shown in Fig. 1a, when the sample is exposed to a perpendicular magnetic field $B$ at 12 mK. The filling factor $v$ is defined as $nh/eB$ (where $n$ is the electron density, $h$ is the Planck constant, and $e$ is the unit charge), which represents the number of occupied Landau levels without spin degeneracy. In high quality samples, electron solid states can form at appropriate filling factors with a magnetic field applied, well known as bubble states [18, 21]. In bubble states, the conductance is featured as dips toward zero (highlighted with green color in Fig. 1a), same as that in the incompressible quantum fluid of fractional quantum Hall (FQH) states [32], such as $v = 8/3$, $v = 5/2$ and $v = 7/3$ states. The Hall resistance and bias breakdown behaviors (see Supplemental Material Section 2) can easily distinguish bubble states from FQH states, which has been well studied in those states [18, 19, 33].

Four bubble states are referred to as R2a, R2b, R2c and R2d (Fig. 1a) respectively as in previous works [21], which are used to study the dynamics of electron solids under external drive. The microscopic nature of the bubble states is understood as the formation of electron bubbles pinned by disorder [11-14], as shown in the left inset of Fig. 1a. Each small dot represents an electron bubble containing multiple electrons, and all of these electron bubbles form a lattice structure due to interaction [11-14]. Meanwhile, there may be distortions and disclinations in the lattice according to numerical simulation results of electron solids [25]. Applying an electric field $E$, realized by adding a dc bias voltage between the inner and outer contacts, depins the localized electron bubbles. In Fig. 1b, the R2c state ($B = 4.85$ T) is shown as an example. The conductance dramatically deviates from zero at a threshold field $E_t = 1.30$ V/m. Further increasing $E$, the general trend of the trace decreases, accompanied by unstable conductance and hysteresis, which represents a partially moving electron solid. After $E$ exceeds the last conductance dip field $E_d = 11.30$ V/m, as shown in Fig. 1b, the conductance trace becomes smooth. $E_t$ and $E_d$ separate the map into three distinct regions, A, B, and C, as noted in Fig. 1b. Similar features can also be observed in R2a, R2b and R2d states (Fig. S2). In contrast to bubble states, the breakdown of integer quantum Hall (IQH) and FQH (Fig. S2) states are much gentler, without fluctuation and hysteresis [19].

The measured conductance signals reflect the overall flow of the carriers while the details of their motion are discarded, and thus the conductance measurement alone does not directly reveal the microscopic dynamic information. Other experimental methods such as surface acoustic wave [34, 35] and resistively detected nuclear magnetic resonance [36] can be used to study the microscopic nature of the electron solids, but they are both focused on electron density distributions and are insensitive to dynamic properties. Indeed, as a charged system, the dynamic ordering of flowing charged units can also be reflected in electrical transport signals. If the depinned charged units flow steadily and have spatial periodicity, such a dynamically ordered structure, when interacting with disorder or the boundary, will cause periodic current fluctuations in the time domain. Corresponding to the time domain signal, there will be peaks at finite frequencies in the spectrum, which is called narrow-band noise [23, 37-39]. Whereas if there is a wide distribution of enhanced noise signal in the frequency domain, the dynamic structure of the system is considered to be disordered [23-25]. This type of noise signal is called broad-band noise here, in order to distinguish it from the narrow-band noise [40-42]. Therefore, noise spectrum measurements can be used to explore the dynamic ordering information of the electron solids.

Both broad-band noise and narrow-band noise exist in our spectrum measurements. The results of the R2c state at 12 mK are shown as an example in Fig. 2, and details of the experimental setup are depicted in Methods and Fig. S1. Noise spectrum in Fig. 2a exhibits a prominent pure narrow-band noise signal generated along the noise floor (dashed blue line) with a fundamental frequency $f_0$, accompanied by harmonics $2f_0$, $3f_0$, $4f_0$, etc., when $E = 6.20$ V/m. This result indicates that there exists one type of dynamically ordered structure in the system. In striking contrast, by changing $E$ to 3.60 V/m (Fig. 2b), noise spectral density $S_I$ enhances from the noise floor (dashed blue line) broadly, especially at the low-frequency range. Although some individual peaks can also be observed, the appearance of the prominent broad-band noise indicates a dynamically disordered structure has formed. Our noise spectra show that different dynamic orderings can emerge in an electron solid, and they are tunable by changing the external drive.

To systematically study the dynamic ordering transition, we probe the evolution of the conductance $G$ and the noise spectra as a function of $E$. The results are shown in Fig. 2c & 2d. Qualitatively, enhanced noise signals are seen in region B (between two dashed black vertical lines). Narrow-band noise characterized by tilted narrow stripes can be recognized clearly. In addition, broad-band noise, which occupies a large frequency band ranging from zero to several kilohertz, appears discretely. The boundaries of the broad-band noise regions are marked by dashed white vertical lines. Remarkably, the phenomenon that broad-band noise appears discretely over the narrow-band noise as $E$ increases indicates dynamically disordered and ordered structures dominate alternately, which has not been predicted before, either theoretically or computationally. Therefore, according to the noise signals in Fig. 2d, we further divide the region B (Fig. 1b) into three types of sub-regions: region $B_0$ shows no noise signal; region $B_1$ and $B_2$ are the regions where there are noise signals, while broad-band noise only emerges in region $B_1$. Such region classifications are also consistent with the conductance features in Fig. 2c: the conductance in region $B_1$ fluctuates more violently than that in region $B_2$. Moreover, the conductance jumps or kinks at the boundaries of region $B_1$ and $B_2$, signifying the dynamic transitions between these regions [23].

The process of dynamic ordering transitions is summarized in Fig. 2e. For $E < E_t$ (1.30 V/m), which is located in region A as classified in Fig. 2c, the conductance is zero, and no noise signal can be observed, indicating that the electron solid remains pinned. When $E$ is slightly greater than $E_t$ (1.30 V/m $\leq E <$ 3.36 V/m, corresponding to region $B_0$ in Fig. 2c), the conductance becomes a finite value, indicating there are movable depinned electron bubbles, but they may be too infrequent to generate observable noise signals. The electron solid lattice might tear, and the depinned electron bubbles move along self-established channels, as predicted by numerical simulations [24, 25]. Further increasing $E$, the structure of the system alternates between dynamically ordered and disordered structures. Finally, for $E > E_d = 11.30$ V/m, when entering region C in Fig. 2c, the conductance tends to be stable, and the noise signals no longer exist even when the value of $E$ is doubled (see Fig. S5), which suggests that the electron bubbles should be destroyed at strong enough $E$. We note that dynamic ordering alternation behaviors can not only be repeated in the same R2c state after magnetic field cycles or temperature thermal cycles, but also appear in other bubble states such as the R2d state (Fig. S3). As a comparison, neither broad-band noise nor narrow-band noise appears in the quantum Hall states, such as in the $\nu = 5/2$ FQH state (Fig. S4) and the $\nu = 3$ IQH state (Fig. S4). The generation and alternation of the dynamically ordered and disordered structures are related to the periodicity status of the electron solids.

To study the stability of the dynamic ordering alternations, temperature dependent noise spectrum measurements were performed for the R2c state. Figure 3a shows the conductance of the R2c state as a function of the temperature. Increasing the temperature, the system becomes electrically conductive from zero conductance at 12 mK. A conductance peak appears around 42 mK, which is the melting temperature $T_{melt}$ of the electron solid, indicating the original insulating electron solid melts into conductive electron liquid, consistent with the previous work [21]. Further increasing the temperature, the conductance drops, and fewer electron solids are expected to exist in the system. Indeed, no noise signal is observed at 45 mK (Fig. 3b and Fig. S6b) or 100 mK (Fig. S6c), confirming the melting scenario. When the temperature is less than $T_{melt}$, and the sample has finite conductance, i.e., between about 25 mK and 42 mK, the system is in a mixed phase with both electron solid and electron liquid. At 35 mK (Fig. 3c) much weaker and less noise signal is observed than that at 12 mK (Fig. 3d), which further supports the existence of the mixed phase. Nevertheless, the dynamic ordering alternations persist even when electron solid is mixed with electron liquid. We note that the measured conductance means the number of charged units that could be transported through the system within unit time under an external voltage. As shown in Fig. 3a, below and above $T_{melt}$, the measured conductance is similar, but the noise signals show dramatic differences, demonstrating that noise spectrum is a powerful tool in studying the charged solids in electronic systems.

## 3. Discussion

Previous numerical simulations on electron solids show that dynamically disordered structures can transit to dynamically ordered structures by increasing the external drive [24]. The lattice of a pinned electron solid can exhibit disclinations if the pinning is strong enough [25], and in this case, only part of the charged units start to move when the driving field reaches the depinning threshold corresponding to plastic depinning transition [24, 25, 43]. The flow of the depinned charged units is initially disordered, characterized by crossing winding channels or percolation-like paths, which is called plastic flow phase [24]. Under high driving field the disordered plastic flow phase can transit to an ordered moving smectic phase [23, 24]. Critical phenomena near the depinning threshold may exist, which take the form of $v_b \sim (F - F_t)^\beta$, where $v_b$ is the sliding velocity of electron solids, $F$ is the driven force, $F_t$ is the depinning threshold force, and $\beta$ is the critical exponent [23, 24]. For plastic depinning transition, the critical exponent $\beta$ is larger than 1, although its specific value measured in different systems could be different [23, 24, 44, 45]. In this work, driving force $F$ and depinning threshold force $F_t$ is proportional to the driving field $E$ and depinning threshold field $E_t$, respectively. $v_b$ is equal to $\lambda f_0$, where $\lambda$ is the periodicity of the electron solid lattice and thereby a constant in a given bubble state [37]. Therefore, the expression for the critical behavior can be transformed into $f_0 \sim (E - E_t)^\beta$. As for $E_t$ at elevated temperatures, for example at $T = 35$ mK, value of $E_t$ cannot be extracted from the depinning conductance trace (Fig. S6a) directly, due to the presence of conductive electron liquid at zero bias. Here we analyze the critical behavior by assuming that $E_t$ (1.30 V/m) does not change with the temperature, and in Supplemental Material Section 5, we demonstrate that the choice of $E_t$ does not affect our conclusion. Main features of the noise signals in Fig. 3c & 3d are summarized in Fig. 3e & 3f, respectively. All these figures are in log-log plots for calculation of $\beta$. The narrow-band noise regions are easy to perform such a calculation because the parallel fundamental frequency $f_0$ as well as its harmonics can be clearly recognized. The dashed purple straight lines are power law fittings of the red stripes, i.e., $f_0 \sim (E - E_t)^\beta$, and their slopes in the log-log plots correspond to the values of $\beta$. We extracted $\beta = 3.29 \pm 0.03$ at 35 mK and $\beta = 2.63 \pm 0.07$ at 12 mK, both larger than 1, suggesting a plastic depinning transition. Moreover, our results show that $\beta$ at 35 mK is larger than that at 12 mK, no matter how $E_t$ was chosen (Supplemental Material Section 5). So far, we are not aware of any theoretical or simulation studies on the temperature effect of the plastic depinning transition and the critical exponent, which remains an open question.

Turning dynamically ordered to disordered structures repeatedly by increasing driving field or the alternation between these two kinds of structures in depinned collectively moving systems have never been predicted or observed before. Considering that depinned electron bubbles can flow like a liquid and host different dynamic structures, we analyze the observations from the perspective of hydromechanics related to the transition to turbulence [46]. At the transition between laminar flow and turbulent flow, a sequence of unstable flows can emerge as the Reynolds number increases under proper conditions, and the noise spectrum of the fluid velocity can be used to distinguish the dynamic features of the flow [46, 47]. A periodic flow, whose velocity oscillates periodically with time, corresponds to a narrow-band noise spectrum, whereas a turbulent flow corresponds to a broad-band noise spectrum. In addition, these two kinds of flows could mix together when there are both broad-band and narrow-band noise in the spectrum [46, 48]. Therefore, that the flow of the charged objects changes from dynamically ordered to disordered structures in this work may share similar process with a turbulent transition. Furthermore, there are also red stripes with narrow-band noise characteristics in the broad-band noise regions in Fig. 3d, so the dynamically disordered structures may belong to a mixed flow. The alternation of dynamic structures may be caused by competition between periodic flow and mixed flow rather than caused by turbulent flow which is generally considered to appear under high drives.

## 4. Conclusion

In summary, we report the observation of field-driven noise spectra in a two-dimensional electron solid, showing the existence of alternating dynamic ordering under external drive. Such an unexpected alternation of structures unveils the dynamics of collective motions that could not be detected through conventional transport measurements. The method in this work has not only been demonstrated to be a powerful tool for exploring the dynamics of electronic systems, but also potentially provided new insights into collective motions of complex systems in nature.

## 5. Methods

### 5.1. Sample information

The 2DEG is 309 nm underneath the surface, located in a symmetrically doped $Al_xGa_{1-x}As/GaAs/Al_xGa_{1-x}As$ quantum well ($x = 0.238$) grown by molecular beam epitaxy method. The quantum well width is 28 nm. The sample was fabricated into a Corbino geometry with an inner (outer) contact diameter of 1.4 mm (1.6 mm) and illuminated by a red light-emitting diode at 4.5 K before measurements. The carrier density is $3.0 \times 10^{11}$ cm$^{-2}$, measured by Shubnikov–de Haas oscillations at base temperature. Well-developed bubble states and even-denominator FQH states in the second Landau level (Fig. 1a) suggest the sample should be of ultra-high quality. The mobility is difficult to be directly extracted from a sample with Corbino geometry because the two-terminal measurement cannot avoid the contact resistance. In a van der Pauw sample from the same wafer with identical bubble states and FQH states, the mobility is $2.8 \times 10^7$ cm$^2$/Vs.

### 5.2. Experimental environment

Conventional transport and noise spectrum measurements were performed in a dilution refrigerator (Leiden Cryogenics B.V. MNK126-450). Thermocoax filters and cryogenic RC filters, which are connected to each lead of the sample, are used to thermalize the electrons of the sample. The electron temperature is equal to the refrigerator temperature above 12 mK (Fig. S11). The base temperature of the fridge is lower than 6 mK based on calibration from Leiden, cross-checked by an in-house cerium magnesium nitrate thermometer.

### 5.3. Conventional transport measurements

The conductance was measured with standard lock-in techniques. An ac excitation (10 μV at 17 Hz) was applied between inner and outer contacts, and the ac current was measured by the ac voltage drop on a 2 kΩ resistor in series with the sample using a lock-in amplifier. The ac and dc voltages were applied together by a waveform generator (Agilent 33220A) when performing depinning and noise spectrum measurements.

### 5.4. Noise spectrum measurements

The noise spectral density was measured by a low-noise current amplifier (NF-CA5350, gain 1MV/A) and then digitized by a data acquisition card (National Instrument USB-6289, 18 bits, sampling rate 600,000 Hz) outside the cryostat at room temperature. The conductance was measured by the same current amplifier and recorded by a lock-in amplifier simultaneously when performing the noise spectrum measurements. A detailed schematic is shown in Fig. S1.

5.5. Heating effect

When performing noise spectrum measurements, high dc voltages dropped on the sample would, nevertheless, produce Joule heat on the sample. However, considering the fridge cooling power (~ 1 μW at 12 mK) at the measured temperature, the Joule heating on the sample (< 67 pW) is negligible.

The estimation of the Joule heating power is performed in the following way. First, the applied ac voltage (10 μV) is much smaller than the maximum dc voltage (2.20 mV), so is negligible in the estimation. In Fig. 2c and Fig. S5a, the dc current at 2.20 mV can be calculated by integrating the (differential) conductance over the dc voltage, and the result is $I_d$ = 30.12 nA. Hence the corresponding heating power is $P_h$ = 66.26 pW.

Even if the heating effect made a slight change on the sample temperature, this would not be the reason for the observed dynamic ordering alternation behavior. On one hand, according to Fig. 3b-d, elevated temperature weakens or eliminates the noise signals instead of generating noise signals. On the other hand, the positive (differential) conductance, as shown in Fig. 2c and Fig. S6a, leads to a monotonic increase in heating power as the driving field increases, and such a monotonic heating behavior is not sufficient to generate an alternating noise signal. Therefore, we can rule out the heating effect for the interesting dynamic ordering alternation behavior.

**Declaration of Competing Interest**

The authors declare that they have no conflict of interest.

**Acknowledgments**

We thank fruitful discussion with C.J. Olson Reichhardt. The work at PKU was supported by Beijing Natural Science Foundation (JQ18002), the NSFC (11921005, 11674009), the National Key Research and Development Program of China (2017YFA0303301) and the Strategic Priority Research Program of Chinese Academy of Sciences (Grant No. XDB28000000). The work at Princeton University was funded by the Gordon and Betty Moore Foundation's EPiQS Initiative, Grant GBMF9615 to L. N. Pfeiffer, and by the National Science Foundation MRSEC grant DMR-1420541.

**Author Contributions**

X.L. and P.W. designed and supervised the project. J.S. and J.N. realized the measurement condition and performed measurements. J.S., J.N., Y.Liu, Y.Li and P.W. analyzed data. L.P. and K.W. provided the 2DEG sample. J.S., X.L. and P.W. wrote the paper with input from all authors.

**FIGURES**

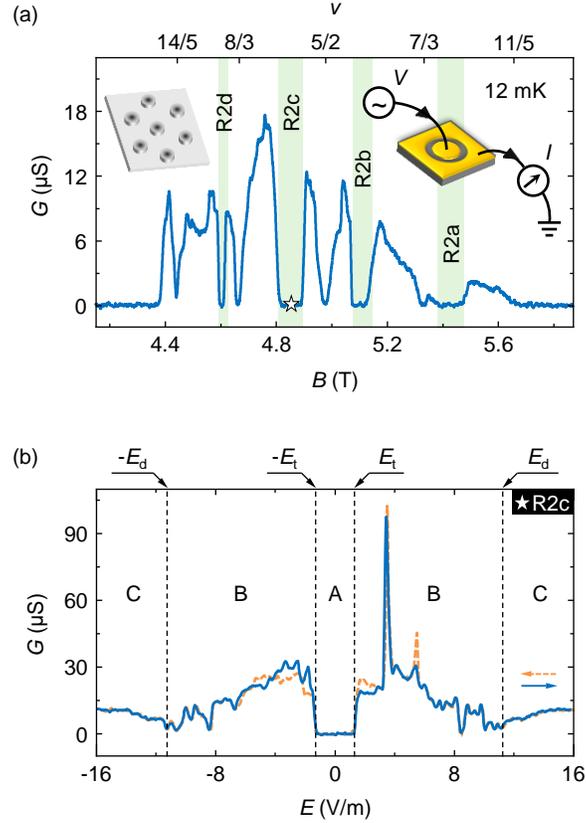

**Fig. 1.** Transport behaviors of bubble states. (a) Conductance $G$ as a function of magnetic field $B$ in the lower spin branch of the second Landau level ($2 < \nu < 3$) at 12 mK. Four green regions indicate bubble states (R2a, R2b, R2c and R2d). The right inset shows the schematic of the conductance measurements. The left inset is a sketch of the bubble states. Each dot represents an electron bubble, and all of them form a triangular lattice. (b) Conductance $G$ as a function of the electric field $E$ in the R2c state ($B = 4.85$ T, the white color star symbol in Fig. 1a) at 12 mK. The $E$ sweeping directions are noted as following: negative to positive (solid blue line) and positive to negative (dashed yellow line). The traces can be divided into three regions by depinning threshold fields $\pm E_t$ and conductance dip fields $\pm E_d$. Region A: zero conductance; Region B: unstable conductance with hysteresis from opposite sweeping directions; Region C: non-zero conductance without obvious fluctuation or hysteresis. For the R2c state, $E_t = 1.30$ V/m and $E_d = 11.30$ V/m.

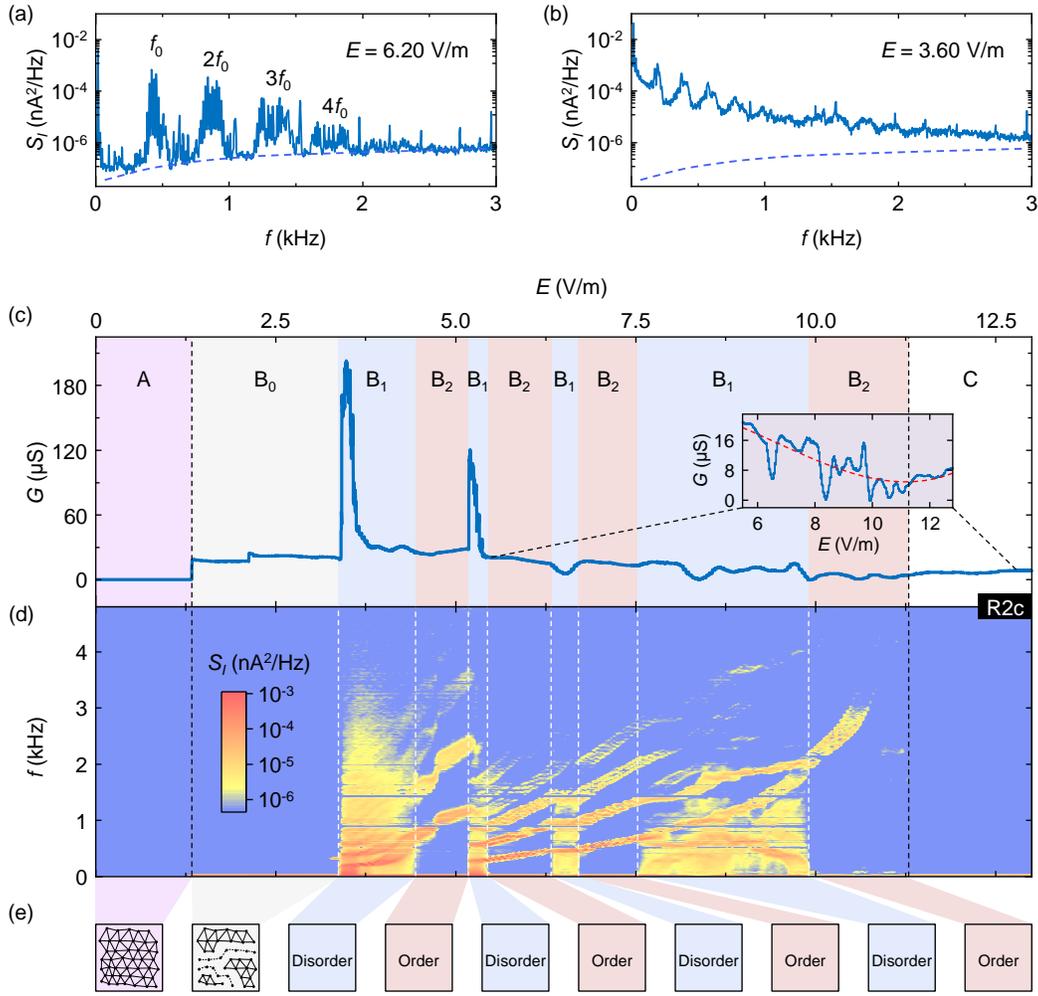

**Fig. 2.** Alternating dynamic ordering of the sliding electron solid in the R2c state. (a) A spectrum dominated by narrow-band noise: a fundamental frequency $f_0$ accompanied by its harmonics $2f_0$, $3f_0$, $4f_0$, etc., generated along the noise floor (dashed blue line). Data were taken at $E = 6.20$ V/m. (b) A spectrum dominated by broad-band noise: the noise signal is dominant for all the frequencies although some individual peaks can be observed. The dashed blue line indicates the noise floor, same as that in Fig. 2a. Data were taken at $E = 3.60$ V/m. (c, d) Conductance $G$ and noise spectra measured simultaneously as $E$ increases. Dashed black vertical lines are $E_t = 1.30$ V/m and $E_d = 11.30$ V/m that separate region A, B and C of Fig. 1b. Inset zooms in the latter part of the conductance trace. The dashed red line is a cubic polynomial fitting of the data to determine the value of $E_d$, whose position is also indicated by the right dashed black vertical line. The blue backgrounds marked by $B_1$ in Fig. 2c are where broad-band noise appears, with boundaries indicated by dashed white vertical lines in Fig. 2d. The red backgrounds marked by $B_2$ in Fig. 2c are where narrow-band noise dominates. Noise spectral density $S_I$ as a function of frequency $f$ and electric field $E$ is plotted with instrumental noise subtracted. Original noise spectra are plotted in Fig. S7. (e) Sketch of dynamic ordering transitions as $E$ increases. The first two boxes represent a global pinned electron solid and a partial depinned electron solid, respectively. The following eight boxes represent the dynamic ordering alternations: the blue (red) shaded boxes represent the dynamically disordered (ordered) structures.

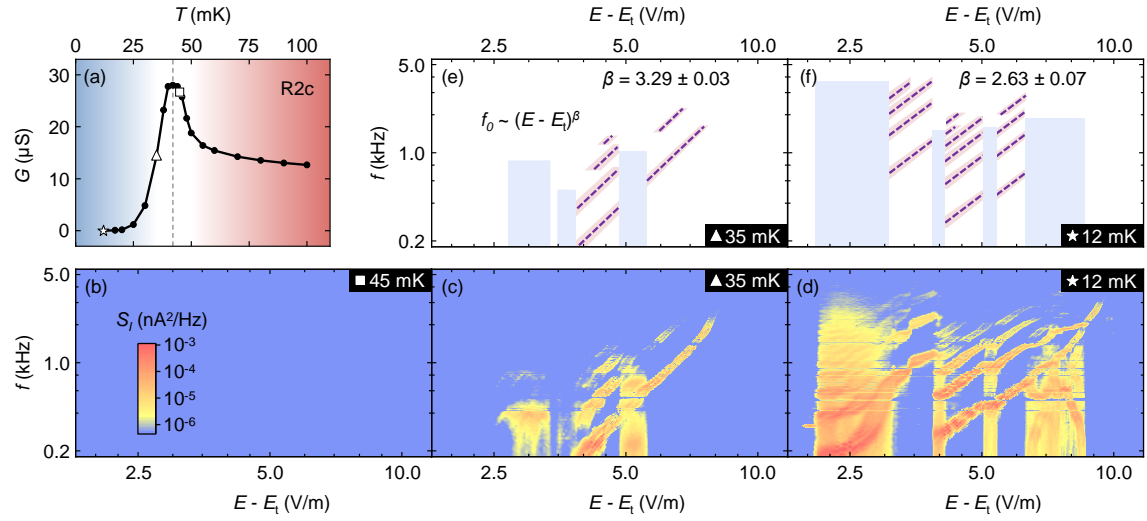

**Fig. 3.** Temperature dependence of the sliding electron solid. (a) Temperature dependence of the conductance $G$ in the R2c state ($B = 4.85$ T). The resolution of $G$ is within $\pm 0.1$ μS. The conductance peak suggests a melting temperature $T_{melt} = 42$ mK (dashed vertical line). (b-d) Noise spectra at 45 mK, 35 mK and 12 mK, respectively. The x-axis is a log scale of $E - E_t$, where $E_t$ is 1.30 V/m. The y-axis is a log scale of $f$. As the temperature increases, the noise signals are weakened and completely smeared out at 45 mK. (e, f) Summarized main features of the narrow-band and the broad-band noise signals shown in Fig. 3c and Fig. 3d, respectively. The blue vertical (red tilted) stripes mimic the areas where broad-band (narrow-band) noise dominates. The dashed purple lines are power law fittings, i.e., $f_0 \sim (E - E_t)^\beta$, of the red stripes, where $\beta$ is the critical exponent.

**Supplemental Material for**

**Dynamic ordering transitions in charged solid**

Jian Sun[1†], Jiasen Niu[1†], Yifan Li[1], Yang Liu[1], L. N. Pfeiffer[2], K. W. West[2], Pengjie Wang[3]*, Xi Lin[1,4,5]*

[1]International Center for Quantum Materials, Peking University, Beijing 100871, China

[2]Department of Electrical Engineering, Princeton University, Princeton, New Jersey 08544, USA

[3]Department of Physics, Princeton University, Princeton, New Jersey 08544, USA

[4]Beijing Academy of Quantum Information Sciences, Beijing 100193, China

[5]CAS Center for Excellence in Topological Quantum Computation, University of Chinese Academy of Sciences, Beijing 100190, China

[†]These authors contributed equally to this work.

*xilin@pku.edu.cn; *pengjie.wang@princeton.edu

**Contents:**

1. Sample geometry and noise spectrum measurement setup

2. Transport behavior of the bubble states and the FQH states

3. Noise spectra for other states

4. Additional experimental results for the R2c state

5. Critical exponent

6. Time-domain signals

7. Electron temperature

## 1. Sample geometry and noise spectrum measurement setup

In a commonly used van der Pauw [1] or Hall-bar [2] shaped 2DEG sample, electron solids form in the bulk, accompanied by quantized edge conductions [3-5]. Therefore, the measured electrical signals contain the information from both the edges and the bulk of the sample. Our sample was fabricated into Corbino geometry [6] (see the sketch in Fig. S1), so only the properties of electron solids were probed without the influence of edge states.

When performing the noise spectrum measurement, the noise and the conductance were measured at the same time. Figure S1 shows the schematic of the measurement setup.

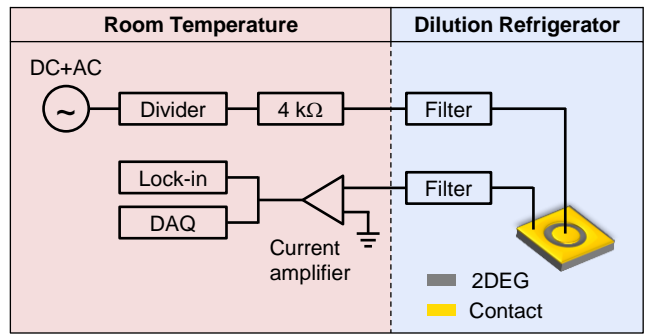

**Fig. S1.** Schematic of the noise spectrum measurement setup. At room temperature, a waveform generator (Agilent 33220A) and a divider were used to apply dc and ac voltages between inner and outer contacts. In the dilution refrigerator, Thermocoax filters [7] and cryogenic RC filters [8] were used to reduce the electron temperature of the sample. The cut-off frequency of the filtering combination was about 16 kHz, which does not influence the noise spectrum measurement. The noise and conductance signals were both amplified by a low noise amplifier (NF-CA5350, gain 1MV/A) outside the cryostat at room temperature. The noise signals were digitized by a data acquisition (DAQ) card (National Instrument USB-6289, 18 bits, sampling rate 600,000 Hz), and the conductance signals were recorded by a lock-in amplifier.

## 2. Transport behavior of the bubble states and the FQH states

Figure S2a-c are depinning traces of the bubble states R2a, R2b and R2d, respectively. Using the same separation criteria as those used in the R2c state, all these traces can also be divided into three distinct regions, A, B, and C, by $E_t$ and $E_d$ (dashed black vertical lines), respectively. Figure S2d-f are the breakdown traces of the FQH states at $v = 8/3$, $v = 5/2$ and $v = 7/3$, respectively. Comparing the trace features of the bubble states and the FQH states, we find that the former ones fluctuate more violently and have pronounced hysteresis from opposite sweeping directions.

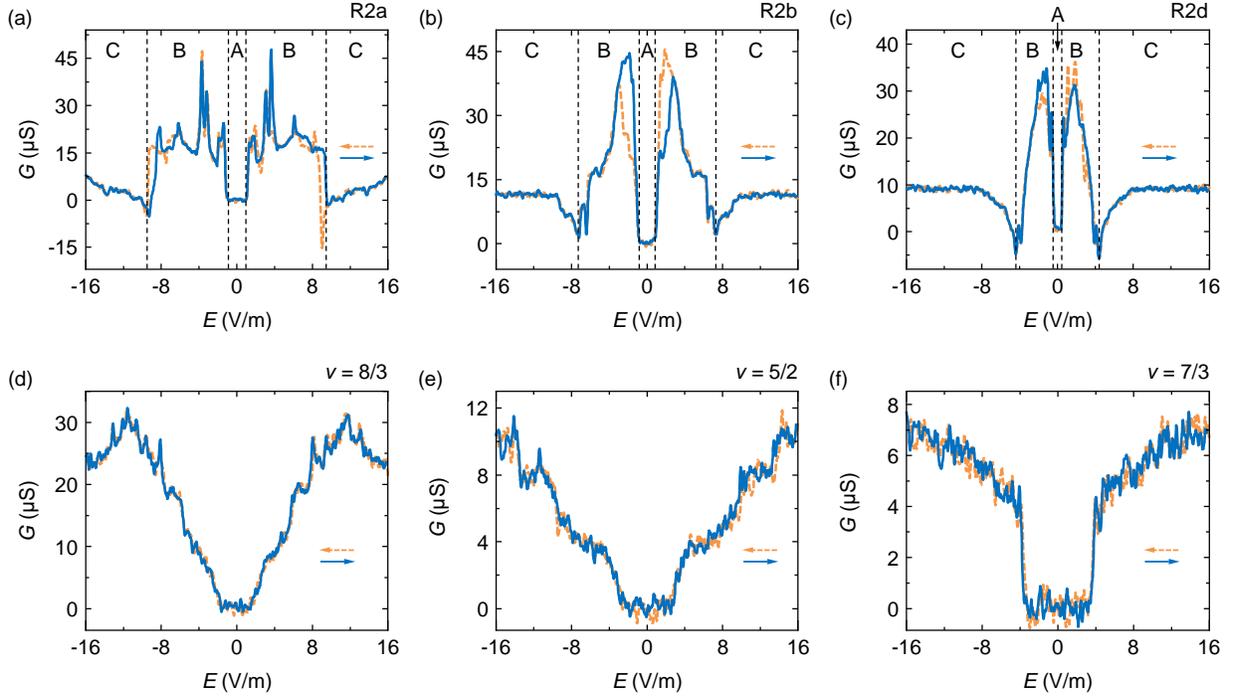

**Fig. S2.** Conductance $G$ vs. electric field $E$ for different states at 12 mK. (a-c) Depinning traces of the bubble states R2a, R2b and R2d, respectively. The dashed black vertical lines are used to separate the traces into three regions: A, B and C. (d-e) Breakdown traces of the FQH states at $v = 8/3$, $v = 5/2$ and $v = 7/3$, respectively. The $E$ sweeping directions are noted as following: negative to positive (solid blue line) and positive to negative (dashed yellow line), as shown in each figure.

## 3. Noise spectra for other states

The dynamic ordering alternation behavior can also be observed in other bubble states, for example, in the R2d state as shown in Fig. S3. The dashed vertical line is a guide line showing where the noise signals disappear, and it is also close to the last obvious conductance dip as shown in Fig. S3a. For comparison, the same measurements were taken in the quantum Hall states, such as in the $v = 3$ IQH state and $v = 5/2$ FQH state (Fig. S4). The results show that no noise signal appears. As for the bubble states in higher Landau levels ($v > 4$), noise spectrum measurements have been carried out in previous works [9, 10], and signs of dynamic ordering alternations can be found in the R4a state from the reference [9].

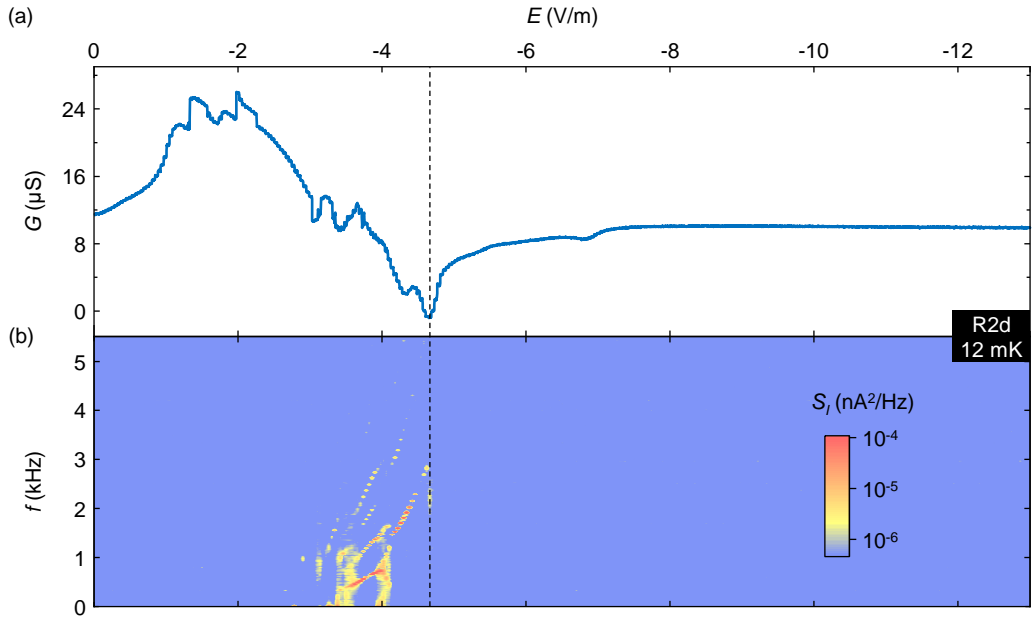

**Fig. S3.** Conductance and noise spectra for the R2d state at 12 mK. (a) Conductance $G$ as a function of electric field $E$. (b) Noise spectral density $S_I$ as a function of frequency $f$ and electric field $E$. The dashed black vertical line is a guide line showing where the noise signals disappear, which is close to the field of the conductance dip. Both Fig. S3a and S3b were measured at $B = 4.62$ T, slightly deviating from the center of the R2d state (4.60 T), and hence the conductance at $E = 0$ V/m has a finite value.

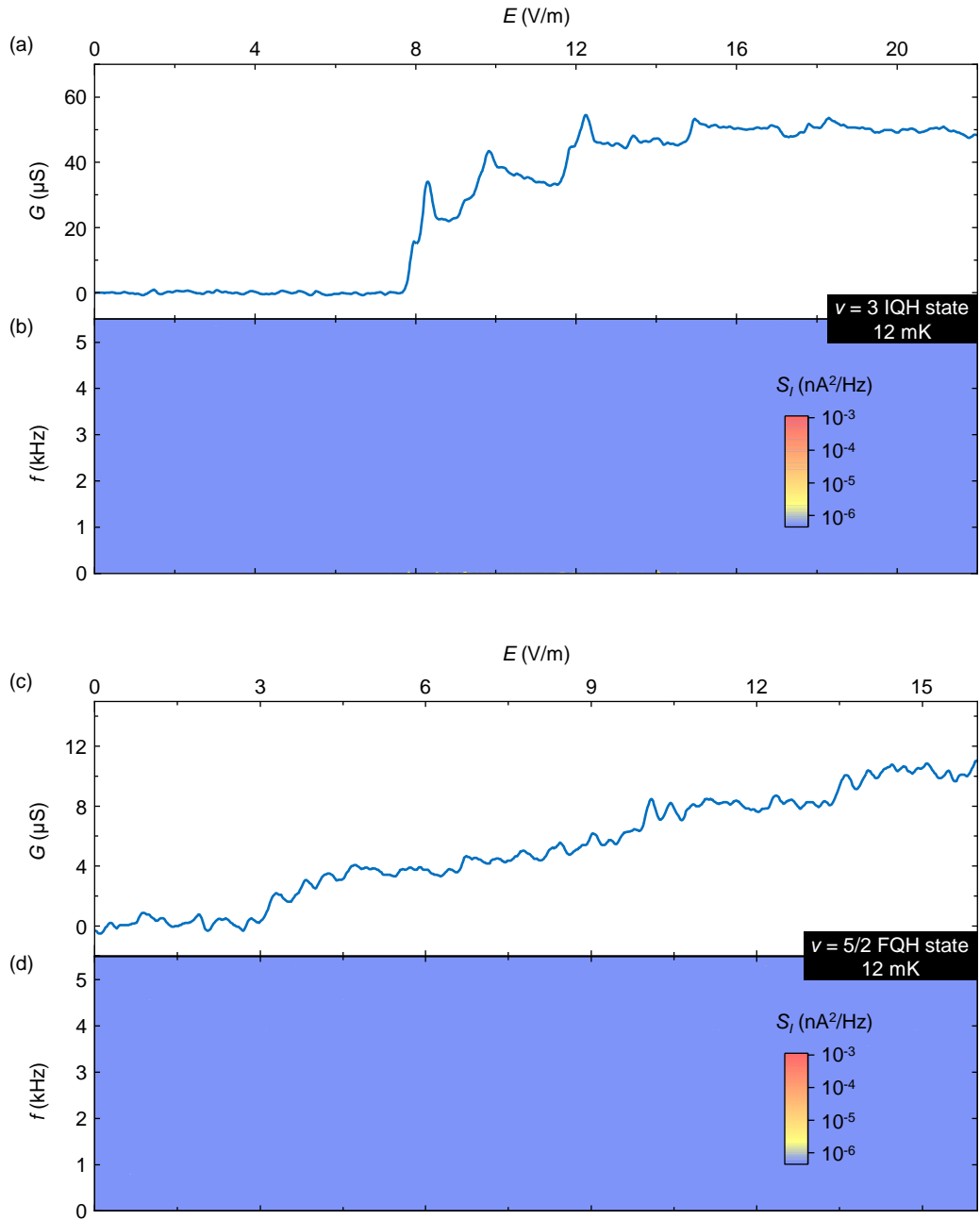

**Fig. S4.** Conductance and noise spectra of the $v = 3$ IQH state and $v = 5/2$ FQH state at 12 mK. There is no narrow-band or broad-band noise in both the $v = 3$ IQH state and $v = 5/2$ FQH state.

## 4. Additional experimental results for the R2c state

To verify that the noise signals do not appear when $E > E_d$ (11.30 V/m), here we present more experimental results for the R2c state at 12 mK with $E$ up to 22.00 V/m (Fig. S5). The results show that narrow-band noise and broad-band noise no longer appear when $E$ is larger than $E_d$. Moreover, temperature dependence of the noise spectra was also measured, and the results are shown in Fig. S6. Similar to the case at 12 mK, both narrow-band noise and broad-band noise disappear when $E$ is larger than the $E_d$ at 35 mK (Fig. S6a). At 45 mK and a higher temperature, 100 mK, no noise signal appears (Fig. S6b & S6c).

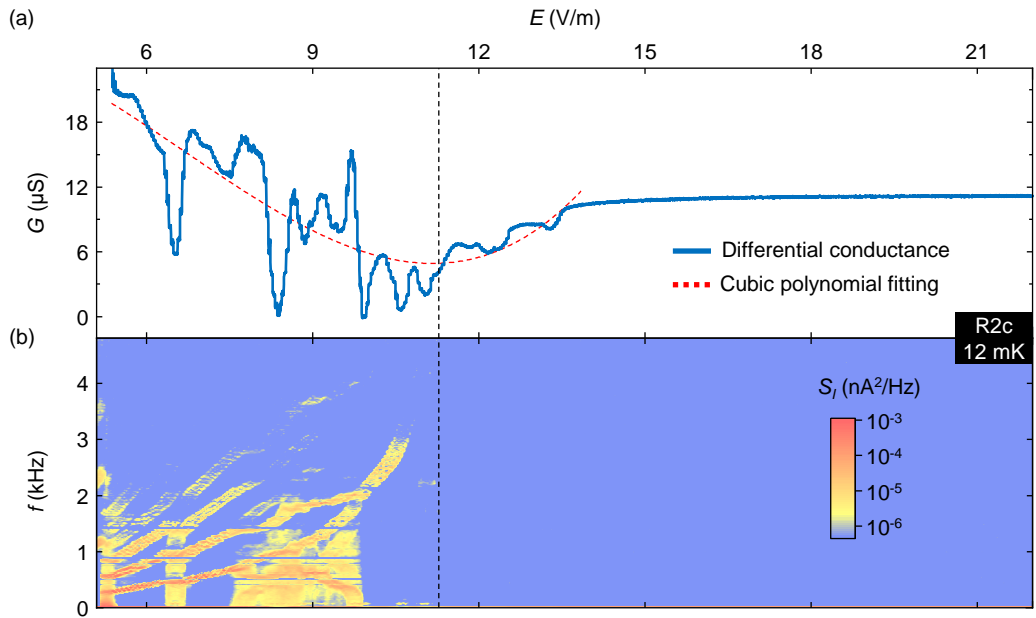

**Fig. S5.** Conductance and noise spectra for the R2c state. (a) Conductance $G$ as a function of electric field $E$ at 12 mK. The dashed red line is the cubic polynomial fitting of the conductance data to determine the value of $E_d = 11.30$ V/m. (b) Noise spectral density $S_I$ as a function of frequency $f$ and electric field $E$ at 12 mK. The dashed black vertical line is a guide line showing where the noise signals disappear, which is near the dip of the dashed red line. There is no noise signal when $E > E_d$.

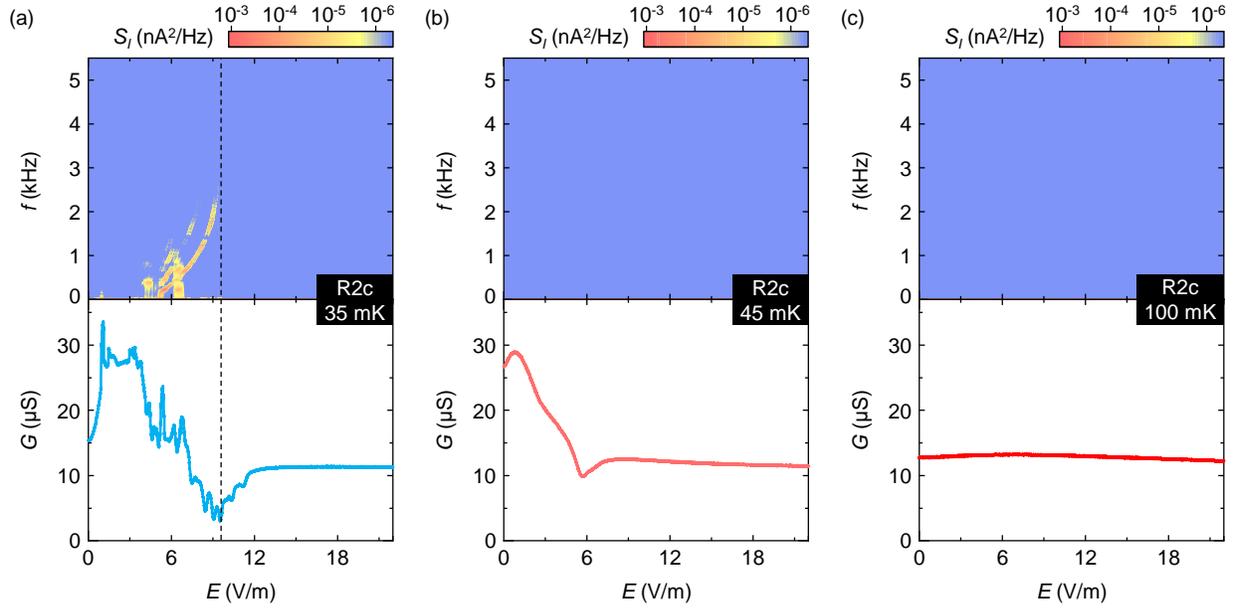

**Fig. S6.** Temperature dependence of the conductance and noise spectra for the R2c state. (a) Conductance $G$ and noise spectra measured simultaneously as a function of $E$ at 35 mK. The dashed black vertical line is a guide line showing where the noise signals disappear, which is near $E_d$. (b, c) Conductance $G$ and noise spectra measured simultaneously as a function of $E$ at 45 mK and 100 mK respectively, and no noise signal appears at these two temperatures.

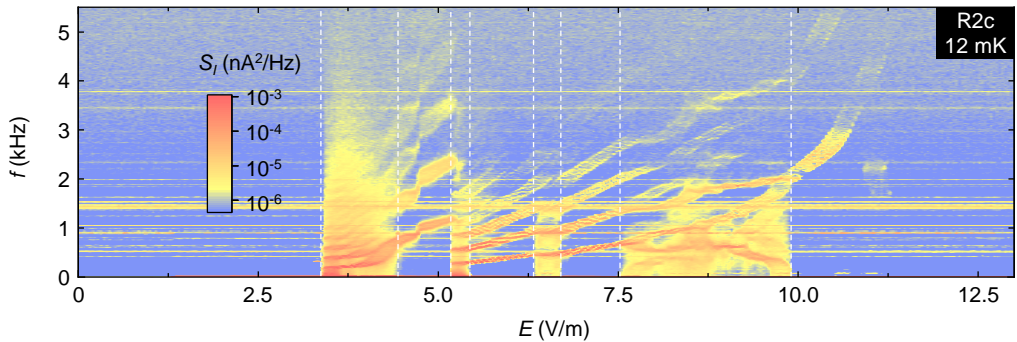

**Fig. S7.** Noise spectra for the R2c state at 12 mK before instrumental noise was subtracted. The horizontal lines running across the entire image are artificial signals from the measurement setup.

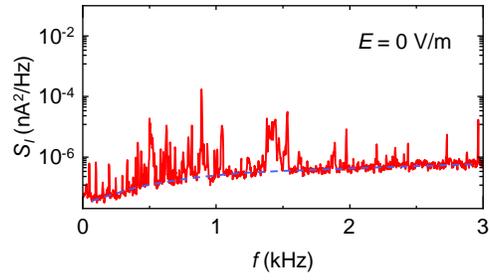

**Fig. S8.** Noise floor. The solid red line is the noise spectrum measured at $E = 0$ V/m, where the electron solid remains pinned, and thus no noise signal from the dynamics is expected to exist. All the sharp peaks are artificial signals from the measurement setup at fixed frequencies, which correspond to the horizontal lines in Fig. S7. The dashed blue line is a guide line indicating the noise floor.

## 5. Critical exponent

In the main text, critical exponent $\beta$ at 35 mK has been calculated by choosing the same $E_t$ (1.30 V/m) as the value used at 12 mK. The results show that $\beta$ at 35 mK is larger than that at 12 mK, and both are larger than 1. The existence of conductive electron liquid under zero bias at 35 mK prevents us to extract the actual value of $E_t$ from the conductance trace (Fig. S6a) directly. Moreover, considering that the thermal motion of electron solids becomes more pronounced at elevated temperatures, the pinned electron solids are easier to be depinned, and thus the actual value of $E_t$ may be smaller than 1.30 V/m. To figure out how $\beta$ changes as $E_t$ decreases at 35 mK, here we choose $E_t = 0$ V/m as an example (Fig. S9). The results show that $\beta$ at 35 mK increases from 3.29 ±0.03 to 4.04 ±0.04 by decreasing $E_t$ from 1.30 V/m to 0 V/m, and thus the conclusion that $\beta$ at 35 mK is larger than that at 12 mK (2.63 ± 0.07) remains unchanged.

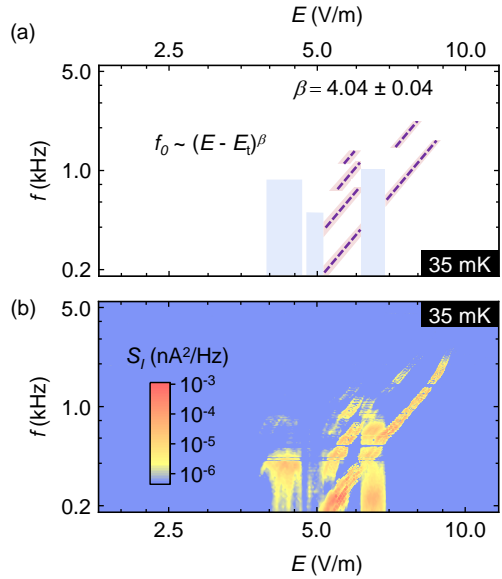

**Fig. S9.** Slope of the narrow-band noise stripes at 35 mK if choosing $E_t = 0$ V/m. The calculated critical exponent $\beta$ changes to 4.04 ±0.04 if choosing $E_t = 0$ V/m.

## 6. Time domain signals

Figure S10a & S10b show the time domain signals corresponding to the spectrum dominated by narrow-band noise (Fig. 2a) and broad-band noise (Fig. 2b) in the main text, respectively. The oscillations of the current background are coming from the external 17 Hz ac excitation. In Fig. S10b, besides the current oscillations, there are also interruptions marked by dashed red circles which contribute to the generation of the broad-band noise in the noise spectrum. The interruptions seem to occur at the bottom of the current background oscillation. To find out whether the generation of the interruptions is related to the 17 Hz ac excitation, we carried out the same measurement without 17 Hz ac excitation. The results show that the interruptions still exist (dashed red circles in Fig. S10c). Therefore, we demonstrate that the generation of the broad-band noise is independent of the 17 Hz ac excitation. The time domain signals in Fig. S10a-c are all digitally filtered (low-pass, from 0 to 5 kHz) to eliminate the influence of high frequency irrelevant noise.

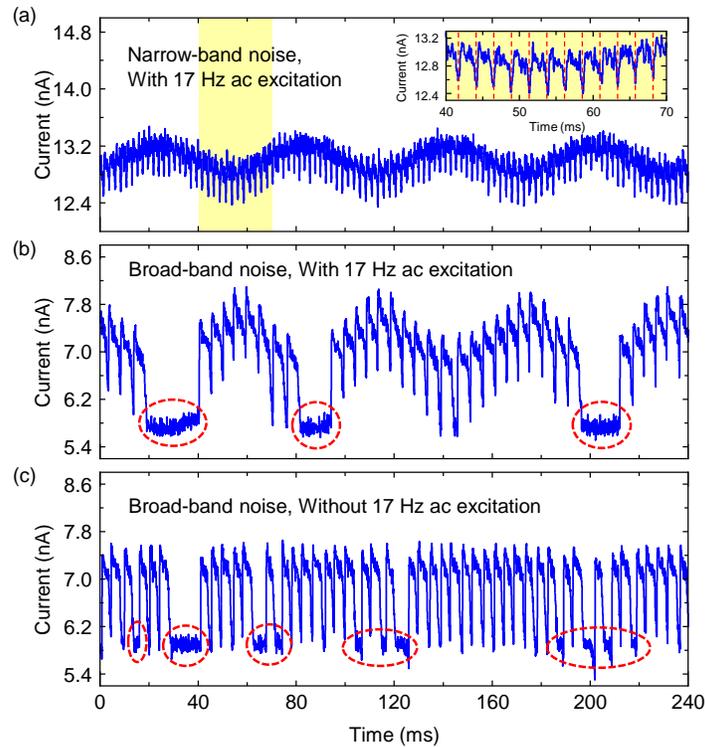

**Fig. S10.** Time domain signal examples for the R2c state. (a) Time dependence of the current signal corresponds to the spectrum dominated by narrow-band noise as shown in Fig. 2a. Evenly-spaced sharp current peaks are the narrow-band noise signals, while the oscillations of the current background are coming from the external 17 Hz ac excitation. The inset shows a zoom-in graph for better clarity of the periodic oscillations, where the evenly-spaced dashed red lines are used for eye-guide. (b) Time dependence of the current signal corresponds to the spectrum dominated by broad-band noise as shown in Fig. 2b. There are several places where oscillations are interrupted (dashed red circles), which generates the broad-band signal in the noise spectrum. (c) Time domain signals without 17 Hz ac excitation were measured in order to exclude the correlation between the interruptions and the 17 Hz ac excitation.

## 7. Electron temperature

The temperature mentioned in the main text is the electron temperature, which is equal to the refrigerator temperatures in this work. The electron temperature is confirmed by the temperature dependence of the $v$ = 14/5 FQH state's conductance as shown in Fig. S11.

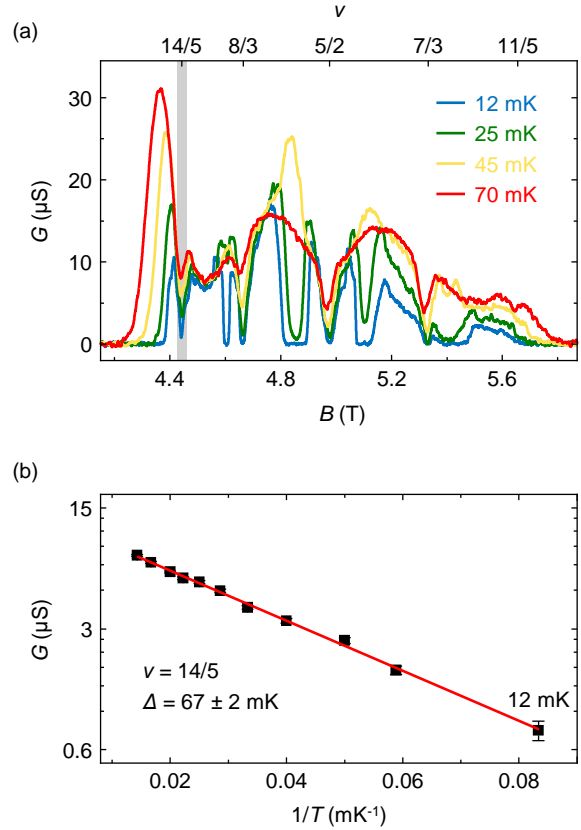

**Fig. S11.** 12 mK electron temperature. (a) Temperature dependence of the conductance in the lower spin branch of the second Landau level ($2 < v < 3$). (b) Arrhenius plot of the 14/5 FQH state (the dip of grey-shaded region in Fig. S11a). The linear relation agrees with that the electron temperature is equal to the refrigerator temperature at least above 12 mK.